# Dual-Tap Optical-Digital Feedforward Equalization Enabling High-Speed Optical Transmission in IM/DD Systems

Yu Guo, Yangbo Wu, Zhao Yang, Lei Xue, Ning Liang, Yang Ren, Zhengrui Tu, Jia Feng, and Qunbi Zhuge*

*Abstract*—Intensity-modulation and direct-detection (IM/DD) transmission is widely adopted for high-speed optical transmission scenarios due to its cost-effectiveness and simplicity. However, as the data rate increases, the fiber chromatic dispersion (CD) would induce a serious power fading effect, and direct detection could generate inter-symbol interference (ISI). Moreover, the ISI becomes more severe with the increase of fiber length, thereby highly restricting the transmission distance of IM/DD systems. This paper proposes a dual-tap optical-digital feedforward equalization (DT-ODFE) scheme, which could effectively compensate for CD-induced power fading while maintaining low cost and simplicity. A theoretical channel response is formulated for IM/DD transmission, incorporating a dual-tap optical equalizer, and the theoretical analysis reveals that for an IM/DD transmission using 1371nm over 10km standard single-mode fiber (SSMF), frequency notch is removed from 33.7GHz to 46GHz. Simulation results show that the DT- ODFE achieves an SNR gain of 2.3dB over IM/DD systems with symbol-space feedforward equalizer (FFE) alone. As the fiber length increases to 15 km, DT-ODFE performs well, while FFE, decision-feedback equalizer (DFE) and Volterra nonlinear equalizers (VNLE) all fail to compensate for the power fading and the 7% hard-decision FEC limit is not satisfied. For 200 Gb/s/λ PAM-4 over 15km SSMF, results show that the signal-to-noise ratio (SNR) of the proposed DT- ODFE with optimal coefficients satisfies the 7% hard-decision FEC limit, which uncovers the great potential of the DT- ODFE for high-speed IM/DD systems in LR/FR scenarios.

## I. INTRODUCTION

IN the era of 5G and beyond, the demand for high-speed data transmission has surged, driven by applications such as artificial intelligence, high-definition video streaming, cloud computing, and virtual reality. According to a recent report from Cisco, 86% of the worldwide Internet traffic is data center interconnection (DCI) related, with 77% of that traffic being within data centers. To address such demand, the evolution of short-reach communication systems, especially within the confines of DCI and access networks, has become paramount [1], [2]. Moreover, as the demand for capacity increases rapidly, 800G/1.6T interconnection schemes and optical components are developed in DCI and access networks such as mobile fronthaul (MFH) [3]. In a survey conducted by Google, all the critical optical and electrical components should be ready in 2023 [4], including InP externally modulated laser (EML), InP and Silicon Photonic (SiP) photodetector (PD) with a 3dB bandwidth of > 55GHz. Hence, high-speed transmission is required for the development of DCI, MFH, and passive optical network (PON). Intensity-modulation and direct-detection (IM/DD) transmission is the most practical solution due to the low complexity and power consumption. There are several fundamental ways to improve the capacity of IM/DD transmission, including higher baud rate, higher order modulation formats, and more parallel lanes, and these axes have been taken the most advantage, from 10Gb/s using 10 Gbaud, PAM-2 and a single lane, to 800Gb/s using 50Gbaud, PAM-4 and 8 lanes [5]. Furthermore, it has been shown that pursuing a higher symbol rate beyond bandwidth limitation is a better choice than using higher-order modulation formats at the expense of a lower symbol rate [6]. However, as the symbol rate increases to 100Gbaud/λ in the 200G era, IM/DD transmission would suffer from more severe impairments, including bandwidth limitation, multipath interference (MPI), and chromatic dispersion (CD). Moreover, CD is one of the essential constraint factors, and direct detection turns CD into power fading in the signal spectrum. The zero-dispersion wavelength (ZDW) window in O-band is being considered for wavelength division multiplexing (WDM) based on 5G MFH to avoid CD-induced power fading [7]. Nevertheless, the performance of WDM transmission using the ZDW is highly limited by the inter-channel four-wave-mixing (FWM).

Many schemes have been proposed to mitigate CD-induced power fading. Dispersion compensating fiber (DCF) was considered a promising way to remove the impact of CD [8]. Two Mach-Zehnder interferometers (MZIs) were used to mitigate the intersymbol interference (ISI) resulting from band-limited transmitters, optical filtering, and CD [9]. In [10], some optical line codes are used to eliminate the effects of CD. Voltage bias tuning in a signal-drive dual parallel Mach-Zehnder modulator (MZM) extends power fading frequency notches at different distances [11]. However, most of the mentioned schemes have certain disadvantages, including difficulty in balancing the DCF compensation performance and modal noise from MPI, focusing on nonreturn-to-zero (NRZ) signals, performance penalty over uncoded transmission with no dispersion, and sophisticated bias control.

On the other hand, digital equalization algorithms for IM/DD systems have been studied widely, including feedforward equalizers (FFE), Volterra nonlinear equalizers (VNLE), decision-feedback equalizers (DFE), and Tomlinson-Harashima precoding (THP) [12]-[15]. In practical realization, most of the equalization in PAM-4 digital signal processing (DSP) is performed by the FFE with 10-30 feedforward taps [16]. However, a survey in [17] shows that the computational complexity of the digital equalization schemes mentioned above is more than 10 times the industry numbers. For example, to transmit a 56Gb/s PAM-4 signal over an 80km link, the number of taps per PAM symbol is 452 with the combination of THP in Tx and VNLE in Rx [15]. Hence, one of the main drawbacks of digital equalization algorithms is the unaffordable computational complexity and high power consumption. Besides, in theory, digital equalization cannot perfectly compensate for the power fading caused by CD, as direct detection can only obtain the intensity information of the signal in IM/DD systems. In contrast, optical equalization can compensate for the CD directly from the optical field. To reduce the complexity of the electrical equalizers, a joint optical-electrical feedforward equalization (OE-FFE) is proposed for the ER/ZR IM/DD systems [18]. However, as CD becomes severe, the emphasis is on how to suppress CD-induced power fading, while the OE-FFE, with a single-tap optical delay line (ODL) in optical equalizer and low-complexity digital equalizer, fails to compensate for the power fading and the IM/DD system breaks down, which would be described in our simulations later.

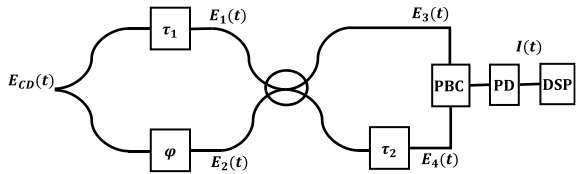

Fig. 1. Schematic diagram of the DT-ODFE.

In this work, we propose a novel dual-tap optical-digital feedforward equalization (DT-ODFE) scheme to suppress severe power fading while reducing the complexity of digital equalizers. Our analysis and experiments focus on LR/FR IM/DD systems, primarily applied in DCI, MFH, and PON. The proposed DT-ODFE consists of two ODLs, a phase shifter before PD and a symbol-spaced feedforward equalizer afterward. The channel response of the IM/DD transmission with two ODLs is described in detail. Compared to the OE-FFE in [18], the proposed DT-ODFE has the advantages as follows: 1)The performance of the optical equalizer is enhanced, and the alleviation of the power fading is more effective. 2)The complexity of the digital equalizer is further reduced. 3)The 3dB power loss of the 1x2 coupler in OE-FFE is avoided by using a 2x2 coupler and a polarization beam combiner (PBC). Thus, a low-complexity digital equalizer suffices to compensate for the residual distortions.

The rest of this paper is organized as follows. In section II, the concept of DT-ODFE is described, and we derive the channel response of an IM/DD system with dual-tap ODLs. The performance of DT-ODFE with different optical coefficients is investigated. In section III, the simulation setup and results are reported.

## II. PRINCIPLE OF DT-ODFE

The schematic diagram of the proposed DT-ODFE is depicted in Fig. 1. The optical equalizer incorporates a 1x2 power splitter, two ODLs, a phase shifter, a 2x2 coupler, and a PBC. We denote the values of two ODLs and the phase shifter as $\tau_1$, $\tau_2$ and $\varphi$ respectively, as shown in Fig. 1. After being transmitted on the fiber, the optical signal suffers from CD and is denoted as $E_{CD}(t)$. The signal $E_{CD}(t)$ is first split into two signals, which pass the first ODL and the phase shifter, denoted as $E_1(t)$ and $E_2(t)$ respectively. Then $E_1(t)$ and $E_2(t)$ are coupled, and the output of the second port of the coupler is delayed by the second ODL. And then $E_3(t)$ and $E_4(t)$ are combined by PBC and directly detected by PD. In Rx DSP, the optical current $I(t)$ is equalized by the digital equalizer of the DT-ODFE. In the following part, we theoretically derive the channel response of an IM/DD system with the optical equalizer and analyze the role of the optical equalizer in mitigating power fading.

### A. Channel Response of the IM/DD System with DT-ODFE

We assume that a small RF signal $cos\omega t$ is transmitted in the IM/DD system, and $\omega$ is the angular frequency. The optical field at Tx output is expressed by $E_{Tx}(t) = e^{j\omega_c t}\sqrt{A + \cos(\omega t)}$, in which the $\omega_c$ is the optical carrier frequency, and A is the direct current (DC) bias [17]. Using Taylor series expansion on optical filed, the envelope of $E_{Tx}(t)$ could be rewritten as [19]:

$$\sqrt{A + \cos(\omega t)} = \sqrt{A} + \frac{\cos(\omega t)}{2A^{\frac{1}{2}}} - \frac{\cos(\omega t)^2}{8A^{\frac{3}{2}}} + \cdots$$
$$\approx \frac{2A + \cos(\omega t)}{2A^{\frac{1}{2}}} = \frac{4A + e^{j\omega t} + e^{-j\omega t}}{4A^{\frac{1}{2}}} \quad (1)$$

The simplified expression includes the optical carrier, an upper sideband, and a lower sideband. A phase shift $\theta$ is induced by CD in the optical field that is correlated with the length of the fiber and the optical wavelength [20]:

$$E_{CD}(t) = e^{j\omega_c t}\frac{4A + e^{j\omega t}e^{j\theta} + e^{-j\omega t}e^{j\theta}}{4\sqrt{A}} \quad (2)$$

in which $\theta = \omega^2\beta_2 L/2$, $\beta_2 = -D\lambda^2/2\pi c$, D is the dispersion coefficient, L is the fiber length, and $\lambda$ is the center wavelength. Then, the signal is fed into the optical equalizer. After passing through the first ODL and the phase shifter, the signals could be written as:

$$E_1(t) = \frac{1}{\sqrt{2}}E_{CD}(t - \tau_1) \quad (3)$$

$$E_2(t) = \frac{1}{\sqrt{2}} e^{j\varphi} E_{CD}(t) \quad (4)$$

We assume the Jones matrix of the coupler is:

$$\begin{bmatrix} 1 & j \\ j & 1 \end{bmatrix} \quad (5)$$

After the coupler, the second ODL would induce a delay $\tau_2$ to the optical signal of the second output port of the coupler. The two signals before the PBC are written as:

$$E_3(t) = E(t - \tau_1) + e^{j(\varphi + \frac{\pi}{2})} E(t) \quad (6)$$

$$E_4(t) = E(t - \tau_1 - \tau_2) + e^{j(\varphi - \frac{\pi}{2})} E(t - \tau_2) \quad (7)$$

The optical field is combined by PBC, which avoids the 3dB power loss compared to the OE-FFE, in which a 1x2 coupler is used in the optical equalizer [18]. After being detected by PD, the DC terms and signal-signal beat interference (SSBI) terms are ignored, and the photocurrent $I(t)$ at angular frequency $\omega$ is:

$$I(t)|_\omega = E_3(t) E_3^*(t) + E_4(t) E_4^*(t)$$
$$= cos\theta cos\omega t + cos\theta cos\omega(t - \tau_1) + cos\omega t cos\left(\theta + \phi + \frac{\pi}{2}\right)$$
$$+ cos\omega(t - \tau_1) cos\left(\theta - \phi - \frac{\pi}{2}\right) + cos\theta cos\omega(t - \tau_2)$$
$$+ cos\theta cos\omega(t - \tau_1 - \tau_2) + cos\omega(t - \tau_2) cos\left(\theta + \phi - \frac{\pi}{2}\right)$$
$$+ cos\omega(t - \tau_1 - \tau_2) cos\left(\theta - \phi + \frac{\pi}{2}\right) \quad (8)$$

According to the relationship between the photocurrent $I(t)$ and the RF signal, the channel response $H(\omega)$ of the IM/DD system with a dual-tap optical equalizer could be denoted as:

$$H(\omega) = cos\theta + e^{-jw\tau_1} cos\theta + cos\left(\theta + \varphi + \frac{\pi}{2}\right)$$
$$+ e^{-jw\tau_1} cos\left(\theta - \varphi - \frac{\pi}{2}\right) + e^{-jw\tau_2} cos\theta + e^{-jw(\tau_1 + \tau_2)} cos\theta$$
$$+ e^{-jw\tau_2} cos\left(\theta + \varphi - \frac{\pi}{2}\right) + e^{-jw(\tau_1 + \tau_2)} cos\left(\theta - \varphi + \frac{\pi}{2}\right) \quad (9)$$

Fig. 2 depicts the channel response over different fiber lengths. When the fiber length is 10km, the wavelength is 1371nm, and the symbol rate is 50Gbaud, the cumulated dispersion is around 21.7ns, and the optical equalizer moved the frequency notch from 33.7GHz to 46GHz, as shown in Fig. 2(a). When the fiber length is 15km, there are two frequency notches in the upper sideband of the channel response for the IM/DD system without an optical equalizer, while only one notch exists in that with an optical equalizer. Besides, it is noticeable that optical equalizers with different parameters led to a different shape of the channel response.

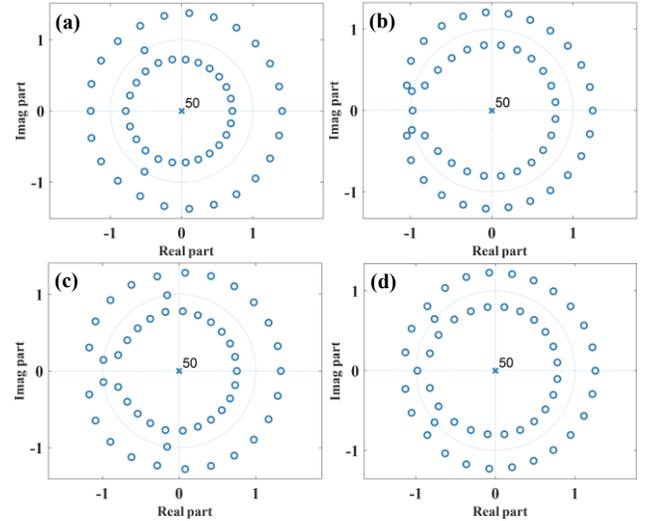

Fig. 3. System pole-zero plots of different distances and optical coefficients of (a)10km, w/o DT-ODFE, (b)10km, $\tau_1$: 14ps; $\tau_2$: 14ps, (c)15km, w/o DT-ODFE, and (d)15km, $\tau_1$: 14ps; $\tau_2$: 14ps.

### B. Performance Analysis of the DT-ODFE

Fig. 3(a)(c) depicts the pole-zero plots of the IM/DD systems without an optical equalizer at different fiber distances, in which "o" and "×" represent the zeros and poles. The zeros on the unit circle correspond to the notches of the channel response in Fig. 2. For instance, there are two zeros on the unit circle in Fig. 3(a), one on the upper part and the other on the lower part, and the two zeros correspond to the frequency notches of the channel response of the IM/DD system without optical equalizer in Fig. 2(a). A digital equalizer in the system without an optical equalizer could not compensate for the CD effectively, resulting from severe noise enhancement at the frequency notches. The optical equalizer makes the notches closer to the point $(-\pi, 0)$, as shown in Fig. 3(b)(d), and the frequency notches are moved to higher frequency points.

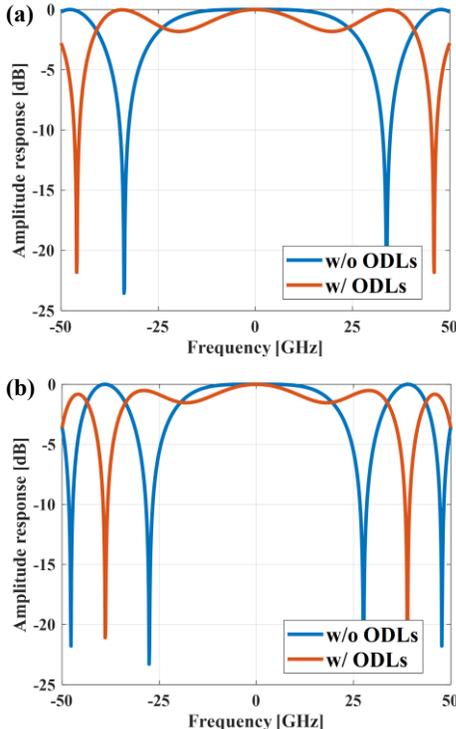

Fig. 2. Channel response of different fiber distances with optical coefficients of (a)10km, $\tau_1$: 14ps; $\tau_2$: 14ps, and (b)15km, $\tau_1$: 14ps; $\tau_2$: 14ps.

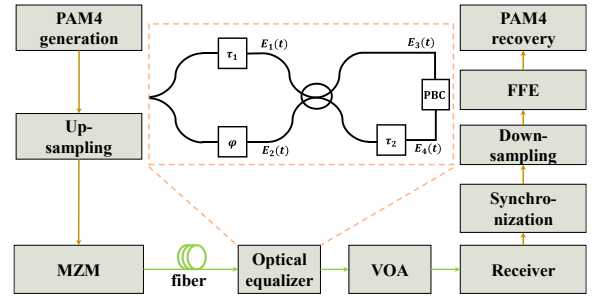

Fig. 4. Simulation schematic diagram of the IM/DD system with DT-ODFE. MZM: Mach Zehnder modulator. VOA: variable optical attenuator.

### III. SIMULATION SETUP AND RESULTS

Based on the theoretical analysis in Section II, we validate the scheme by simulations. WDM transmission in 1.3 $\mu m$ window is one of the most attractive technique for academic research and industrial application, and 1371nm is the edge wavelength of the WDM window [21]. Hence, we set the wavelength in simulations to 1371nm, and the corresponding dispersion coefficient is 7ps/nm/km. The simulation schematic diagram is shown in Fig. 4. In the transmitter, 100000 PAM-4 symbols are generated in MATLAB and resampled to 32 samples per symbol (sps) so that the interval between samples is less than 1ps. The samples are intensity-modulated using an MZM, of which the half-wave and bias voltage are 4V and 3V, respectively. Before intensity modulation, the peak-to-peak voltage (Vpp) of the samples is scaled to 400mV. The samples are transmitted on SSMF, and a variable optical attenuator (VOA) is placed before the receiver to adjust the received optical power (ROP). Before direct detection in the receiver, the samples are equalized by the dual-tap ODLs. In Rx DSP, the samples are synchronized and resampled to 1 sps, and a 15-tap symbol-spaced FFE is applied to the symbols. The first 2000 symbols serve as training symbols, and the taps of the digital equalizer are updated based on the least mean square (LMS).

#### A. Coefficients optimization

The optical equalizer consists of three main optical components: two ODLs and a phase shifter. Hence, there are three coefficients, including the $\tau_1$ and $\tau_2$ for the ODLs, and $\varphi$ for the phase shifter. Different coefficients have a varied effect on the power fading. As shown in Fig. 5, when $\tau_1$ and $\tau_2$ are fixed, DT-ODFE with different $\varphi$ ranging from $-\pi$ to $\pi$ are distinguished in the equalization performance. It could be observed that the optimal $\varphi$ is $-\pi$ or $\pi$. According to Eq. (9), when $\varphi$ is $-\pi$ or $\pi$, the channel response is denoted as Eq. (10); when $\varphi$ is 0, the channel response is denoted as Eq. (11):

$$H(\omega) = (cos\theta + sin\theta) + e^{-jw\tau_1}(cos\theta - sin\theta)$$
$$+e^{-jw\tau_2}(cos\theta - sin\theta) + e^{-jw(\tau_1+\tau_2)}(cos\theta - sin\theta) \quad (10)$$

$$H(\omega) = (cos\theta - sin\theta) + e^{-jw\tau_1}(cos\theta + sin\theta)$$
$$+e^{-jw\tau_2}(cos\theta + sin\theta) + e^{-jw(\tau_1+\tau_2)}(cos\theta + sin\theta) \quad (11)$$

When $\tau_1$ and $\tau_2$ are (14ps, 14ps), the first frequency notch of Eq. (10) is ±38.9GHz, while the first frequency notch of Eq. (11) is ±18.6GHz. Consequently, the optimal coefficient of $\varphi$ is $-\pi$ or $\pi$.

Except for the value of the phase shifter, the coefficients of the two ODLs are also of great significance for the performance of the DT-ODFE. A simulation was conducted to transmit a

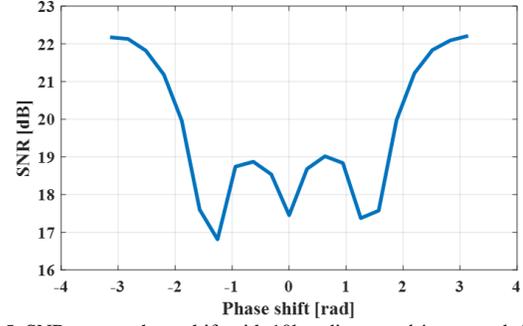

Fig. 5. SNR versus phase shift with 10km distance, 14ps $\tau_1$ and 14ps $\tau_2$.

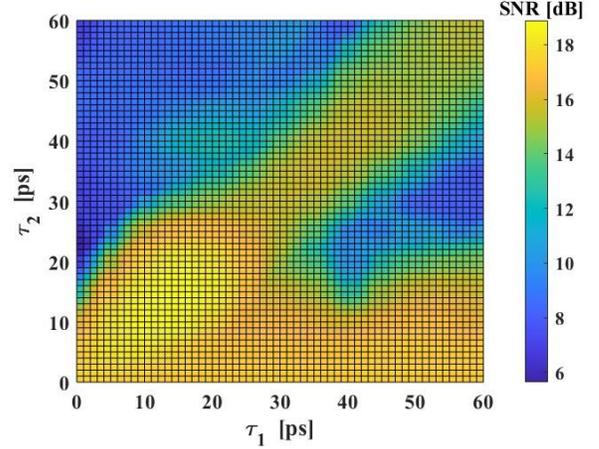

Fig. 6. 2D-optimization results of the delays of the DT-ODFE with a 10km distance and 50Gbaud signal.

50Gbaud PAM-4 signal over 10km SSMF, and the result is presented in Fig. 6. It could be observed that the optimal sets of $\tau_1$ and $\tau_2$ are located around the point of (13ps, 13ps). The coefficients of (0ps, 0ps) indicate that only a symbols-spaced FFE is used in the DSP, and DT-ODFE with the optimal coefficients attains a 1.5dB SNR gain over FFE. In the following discussion, we choose (14ps, 14ps) and (10ps, 6ps) as the coefficients for the ODLs.

#### B. Simulation performance of the DT-ODFE

In Fig. 7, SNR is plotted as the function of fiber distance and ROP, respectively. When fiber is of short length, CD has a negligible impact on the signal. The SNRs of DT-ODFE and FFE all satisfy the 7% hard-decision FEC limitation. As the fiber distance increases and CD impairs the signal severely, FFE cannot compensate for the power fading, and performance degrades rapidly. Contrastingly, DT-ODFE proves adequate in mitigating CD-induced power fading. Significant SNR improvements, surpassing 5dB, are achieved for DT-ODFE of (14ps, 14ps) and (10ps, 6ps) when the fiber length exceeds 13km. Besides, we extensively investigate the compensation effect of DFE and VNLE, as shown in Fig. 7. The performance of the DFE with 15 forward taps and 3 feedback taps and the third-order VNLE is similar, with marginal advantages over FFE. However, the performance of DFE and VNLE are unavoidably impaired by CD-induced power fading, and when fiber length is 14km, the SNR of VNLE cannot satisfy the 7% hard-FEC limit anymore. Noticeably, the corresponding SNR of 7% hard-FEC limit ($3.8 \times 10^{-3}$) is around 16dB for PAM-4 in the additive white Gaussian noise (AWGN) channel. The performance of the DT-ODFE with varied coefficients differs, while both performances satisfy the FEC limit in LR/FR scenarios.

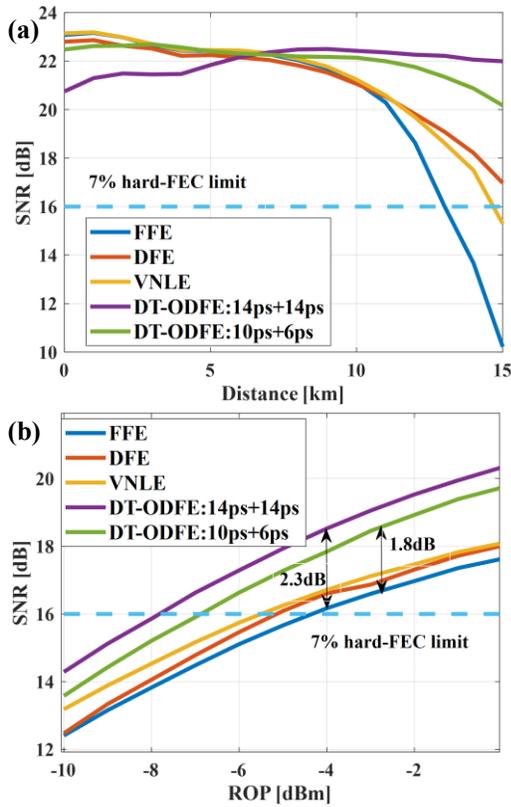

Fig. 7. Performance of the DT-ODFE in (a) varied fiber distance and a fixed ROP of -6dBm, and (b) varied ROP at a fixed 10km distance.

When fiber length is fixed at 10km, ROP is adjusted to evaluate the performance of the DT-ODFE in Fig. 7(b). Results show that the proposed DT-ODFE has a stable SNR gain of around 2.3dB and 1.8dB when ROP ranges from -10dBm to 0dBm. As a comparison, a simulation is conducted to investigate the potential performance of DT-ODFE and OE-FFE at different symbol rates. As shown in Fig. 8(a), when the symbol rate is 50Gbaud, DT-ODFE functions well, but OE-FFE cannot satisfy the 7% hard-FEC limit once the fiber length exceeds 12km. As the symbols rate increases to 100Gbaud in Fig. 8(b), the performance of OE-FFE breaks down while the DT-ODFE still compensates for the CD-induced power fading effectively.

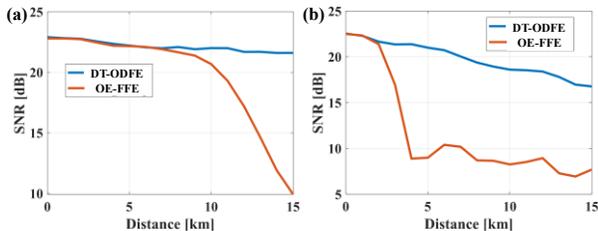

Fig. 8. Performance comparison between DT-ODFE and OE-FFE of (a)50 Gbaud and (b)100 Gbaud.

According to the simulation results, conclusions are listed as follows: 1)Coefficients of the optical equalizer should be optimized. The optical equalizer with higher delays achieves a more considerable SNR gain when severe CD occurs. 2)When fiber length is fixed, the proposed DT-ODFE has stable SNR gain at different ROPs. 3)DT-ODFE enables high-speed transmission with a low-complexity digital equalizer DSP. In a word, DT-ODFE is superior to other digital equalization schemes for its effective compensation performance and low-complexity design. Besides, WDM is a fundamental approach for expanding the capacity of IM/DD systems for DCI, MFH, and PON [22]-[24]. In [25], it has been demonstrated that the single-tap OE-FFE is compatible with dense WDM, and we expect that DT-ODFE also has excellent potential in WDM systems, and corresponding research would be a promising work.

IV. CONCLUSION

In this work, we propose the DT-ODFE scheme, aiming to mitigate the CD-induced power fading in IM/DD transmission effectively with low-complexity DSP. In the DT-ODFE, a dual-tap optical equalizer is used before PD, followed by a low-complexity 15-tap digital FFE. A theoretical analysis of the channel response of the IM/DD transmission with an optical equalizer is formulated in detail. The analytical results indicate that the optical equalizer efficiently eliminates deep notches or moves the notches to higher frequency points of the signal spectrum. We validate the DT-ODFE by simulations, and for comparison, the performance of the DT-ODFE and OE-FFE is also investigated. Results show that the DT-ODFE is more useful in mitigating the CD-induced power fading and has stronger robustness when fiber length and symbol rate increase, especially when the symbol rate is up to 100 Gbaud, which shows great potential for beyond-200-Gb/s/λ optical transmission.